\documentclass[manuscrip]{aastex}

\begin{document}

\title{New high-resolution radio observations of the SNR CTB 80}

\author{G. Castelletti\altaffilmark{1,2}, 
G. Dubner\altaffilmark{1,3},
K. Golap\altaffilmark{4},
W. M. Goss\altaffilmark{4},
P. F. Vel\'azquez\altaffilmark{5}, 
M. Holdaway\altaffilmark{6} and 
A. Pramesh Rao\altaffilmark{7}}

\altaffiltext{1}{Instituto de Astronom\'{\i}a y F\'{\i}sica del Espacio (CONICET, UBA),
C.C.67, Suc. 28, 1428 Buenos Aires, Argentina;
gcastell@iafe.uba.ar; gdubner@iafe.uba.ar}
\altaffiltext{2}{Fellow of CONICET, Argentina}
\altaffiltext{3}{Member of the Carrera del Investigador Cient\'\i fico of CONICET, Argentina}
\altaffiltext{4}{National Radio Astronomy Observatory, 
P.O. Box 0, Socorro, New Mexico 87801; kgolap@nrao.edu; mgoss@nrao.edu}
\altaffiltext{5}{Instituto de Ciencias Nucleares,  UNAM, 
Ap. Postal 70-543, CP. 04510, Mexico City, Mexico; pablo@nuclecu.unam.mx}
\altaffiltext{6}{National Radio Astronomy Observatory, Bulding 65,
949 North Cherry Av., Tucson, AZ 85721-0655; mholdawa@nrao.edu}
\altaffiltext{7}{National Center for Radio Astrophysics, Ganeshkhind, Pune University Campus, 
Pune 411 007, India; pramesh@ncra.tifr.res.in}

\begin{abstract}

We report new high resolution and high sensitivity radio observations of the extended supernova 
remnant (SNR) CTB 80 
(G69.0+2.7) at 240 MHz, 324 MHz, 618 MHz, and 1380 MHz. 
The imaging of CTB 80 at 240 MHz  
and 618 MHz 
was performed using the Giant Metrewave Radio Telescope (GMRT) in India. The observations at 
324 MHz and 1380 MHz were obtained using the Very Large Array (VLA, NRAO) 
in its C and D configurations.
The new radio images reveal faint extensions for the asymmetric arms of CTB 80.
The arms are irregular with filaments and clumps of size 1$^{\prime}$ (or 0.6 pc 
at a distance of 2 kpc). 
The radio image at 1380 MHz is compared with IR and optical emission.
The correspondence IR/radio is excellent along the N arm of CTB 80. 
Ionized gas observed in the [SII] line perfectly matches the W and N edges of
CTB 80.
The central nebula associated with the pulsar PSR B1951+32 was investigated
with an angular resolution of 10$^{\prime\prime}$ $\times$ 6$^{\prime\prime}$. 
The new radio image obtained at 618 MHz shows with superb detail 
structures in the 8$^{\prime}$ $\times$ 4$^{\prime}$ E-W ``plateau'' nebula that hosts the
pulsar on its western extreme. 
A twisted filament, about 6$^{\prime}$ in extent ($\sim$ 3.5 pc), trails behind the pulsar in an 
approximate W-E direction. In the bright 
``core'' nebula (size $\sim$ 45$^{\prime\prime}$),
located to the W of the plateau, the
images show a distortion in the morphology 
towards the W; this feature corresponds to the direction in which the pulsar 
escapes from the 
SNR with a velocity of $\sim$240 km s$^{-1}$.
Based on the new observations,  
the energetics of the SNR and of the PWN 
are investigated.

\end{abstract}

 
\section{Introduction}

The morphology and brightness distribution in supernova remnants, as observed in the 
different
spectral regimes, 
are due to internal and external factors.
The former include the explosion mechanism itself as well as the possible presence
of a neutron star injecting relativistic charged particles during the lifetime of the SNR.
Externally, inhomogeneities in the distribution of the surrounding interstellar matter 
(dense clouds or cavities) 
can modify the expansion of the SN blast wave.

Based on their appearance in the radio domain, SNRs have been classically classified into three
broad categories: shell-like (where electrons are  accelerated at the shock front), 
filled-center 
or plerions (where relativistic particles are provided by an active pulsar) and composite or 
hybrid 
(that combine both characteristics, a hollow shell and central emission). 
High resolution and sensitive radio observations have revealed, however, other
interesting morphological patterns 
such as helicoidal filaments (Dubner et al. 1998),
``barrels'', bi-lobed SNRs, jets components
(Gaensler 1999), etc. 
Furthermore, the availability of sensitive and 
high-resolution X-ray observations 
in recent
years has extended this broad classification 
to include the so
called ``X-ray composite'', i.e. SNRs consisting of a radio shell with centrally peaked X-ray 
emission.
Investigations of brightness distribution and spatial spectral variation along with
multiwavelength comparisons (e.g. radio images with optical, infrared and X-ray ones) are important
to understand the nature and evolution of SNRs.
In addition, 
the study of the
surrounding matter is very useful to disentangle the different physical mechanisms at play.

CTB 80 is an example of a peculiar morphology that does not fit standard 
classifications and whose nature as a SNR is still questioned (Green 2001).
This source, located at a distance of $\sim$2 kpc (Strom \& Stappers 2000),
has a large angular size (over 1$^{\circ}$ from the northern extreme to the 
southernmost limit). It is generally faint, except for the 
central region, where three extended arms overlap in a flat spectrum nebula 
(with spectral index $\alpha$  $\sim$ -0.3, 
where $S_{\nu}$ $\propto$ $\nu^{\alpha}$, and about $\sim$ $10^{\prime}\times 6^\prime$  in size). 
Strom (1987) reported the existence of a compact radio source immersed in this nebula. Later,
Kulkarni et al. (1988) confirmed that the point source was a fast spinning pulsar,
PSR B1951+32.
Different radio observations (Angerhofer et al. 1981,
Strom, Angerhofer, \& Dickel 1984, Mantovani et al. 1985, Strom 1987, Strom \& Stappers 2000) 
show that the pulsar is located within a small
($\sim$ $45^{\prime\prime}$) and flat spectrum component ($\alpha$ $\sim$ 0.0) 
which is called the ``core''. The ``core'' is in 
turn placed 
on the western end of the central nebula, called by Angerhofer et al. (1981) 
the ``plateau'' region.  The three outer arms about 
$30^{\prime}$ long
each,  point to the E, N and SW of the plateau and have
a steeper spectral index ($\alpha$ $\sim$ -0.7).
Over a large part of the remnant there is a high degree of linear
polarization, that rises 
up to 10-15 $\%$
on the central component (Velusamy, Kundu, \& Becker 1976).

The associated pulsar PSR B1951+32 has a period 
of $\sim$ 40 ms
and is moving
toward the SW with the relatively high velocity of $\sim$ 240 
km s$^{-1}$ (Migliazzo et al. 2002). Based on proper motion measurements
carried out by these authors, 
the pulsar's age is 
estimated at $\sim$ 64 kyr.
Gamma-ray pulsations from PSR B1951+32 (E $\ge$100 MeV) have been 
found by Ramanamurthy et
al. (1995) with the EGRET instrument. Pulsed X-rays from this object have 
been reported by \"{O}gelman \& Buccheri (1987), Lingxiang et al. (1993),
Safi-Harb, \"{O}gelman, \& Finley (1995), and Chang \& Ho (1997)
from ${\textit EXOSAT}$, ${\textit Einstein}$, 
${\textit ROSAT}$, and ${\textit RXTE}$ 
observations, respectively. 

X-ray radiation has been imaged from the central region of CTB 80 
using the ${\textit ROSAT}$ and ${\textit Einstein}$ instruments. 
The ${\textit ROSAT}$ image shows emission 
from 
the bright ``core'' and the diffuse $8^{\prime}$ nebula 
E of the pulsar,
similar to the radio emission morphology in the center of the source (Safi-Harb et al. 1995).

In the optical regime, the characteristic structure of this remnant arises from 
forbidden-line emission such as [NII], [SII], [OIII], and the H$_{\alpha}$ line.
The emission in these lines
essentially delineates the radio ``core'' component.
Studies focused on this area reveal filamentary H$_{\alpha}$, [NII], and  [SII] emission, 
extending in an irregular fashion in the E-W direction 
(Angerhofer, Wilson, \& Mould 1980). In the [OIII]
line, the central component itself is observed as a system of filaments showing a shell-type
structure, which brightens in the vicinity of the pulsar (Hester \& Kulkarni 1989). 
A larger area has been recently observed by
Mavromatakis et al. (2001) in H$_{\alpha}$+[NII], [SII],
[OIII] and  [OII] lines.
Both filamentary and diffuse structures are observed in the optical 
domain to the S, 
S-E, S-W, and N of
PSR B1951+32. 

The current work presents high resolution and high sensitivity
radio observations of 
CTB 80 at 240 MHz, 324.5 MHz, 618 MHz, and 1380 MHz of the
entire $\simeq$ 1$^\circ$ source.
The radio data presented here at 240 MHz and 618 MHz were acquired with the
Giant Metrewave Radio Telescope (GMRT\footnote{ The GMRT is run by the National Centre
for Radio Astrophysics of the Tata Institute of Fundamental Research.}) 
located near Pune, in India, 
while the 324.5 MHz and 1380 MHz data have been obtained with the 
Very Large Array of the National Radio Astronomy Observatory (VLA\footnote{The Very Large Array 
of the National Radio Astronomy Observatory
is a facility of the National Science Foundation operated under cooperative
agreement by Associated Universities, Inc.}) 
in its C and D configurations.

\section{Observations and data reduction}

\subsection{Radio continuum at 240 MHz}
 
CTB 80 was observed at 240 MHz with the GMRT (Swarup et al. 1991)
on 2002 May 6 and 7 for a total of 10 hours. 
The source was observed 
with a bandwidth of 8 MHz split into 128 channels.
The relevant flux density scale of our observations was determined using observations of 
3C 286 and 3C 48.
The source 1924+334 was used as the phase and amplitude calibrator.
The observations were carried out in continuous
cycles of 10 minutes on 1924+334 and 30 minutes on CTB 80. The flux density
calibrators, which were also the bandpass calibrators, were observed once for
a total of 45 minutes.
The relevant observational parameters for all observed frequencies are listed in Table 1.

The 240 MHz data  
were fully reduced using AIPS++. To flag the data we used
the ``autoflag'' tool,  searching for interference points across channels and time 
respectively for each
baseline. 
The interference rejection on the data was $\sim$ 15 $\%$.
The data were then bandpass calibrated and combined in channels.

To image CTB 80 we have used wide-field imaging techniques based on a 
multi-facet method (Cornwell \& Perley 1992)
to deal with the non-coplanarity of the visibilities. In order to recover 
the image of the extended CTB 80 remnant, the 240 MHz GMRT data were imaged and self-calibrated 
as a patchwork of
26 subfields; 25 facets on CTB 80 and one outlier field on Cygnus A
(located at about
$8^\circ$ to the NE of CTB 80). 
The imaging process included two-stages. First, a multiscale clean was 
carried out at the center of the image. 
Then, this model
was used as the starting point for the multi-facet wide-field imaging process.
We found that performing the
deconvolution in a two stage scheme, reduces the CLEAN striping
problem.
The resulting image at 240 MHz has an angular resolution of 
26$^{\prime\prime}$ $\times$ 17$^{\prime\prime}$, P.A. =
77$^{\circ}$. For presentation purposes the image displayed in Figure 1 has been convolved with a beam
1.4 times larger than the naturally weighted point spread function obtained.
The final rms noise is 4.6 mJy beam$^{-1}$.

\subsection{Radio continuum at 324 MHz}

The $\lambda$= 90 cm image presented here has been produced
after merging  data taken with the VLA in the 
C and D configurations. 
The C-array data were taken 
on 2000 March 21,
while the shortest 
spacings information was available from the VLA D-array
on 2000 August 14.  
In both cases 26 antennas were available, providing $\sim$ 325
instantaneous baselines. For each array the observations were
made with two 3 MHz bands centered at 321.5 MHz and 327.5 MHz. Each 3
MHz bandwidth was divided into 31 channels. The main purpose of
observing in spectral mode was to  assure better interference editing
capability and to reduce the effect from bandwidth smearing in the
longer baselines.

For both C and D arrays, measurements were interleaved with the
primary flux density calibrators 3C 48 
and 3C 286. 
The latter was also used as bandpass calibrator. Also, regular
observations of the secondary calibrators, 1859+129 
and
2038+513 
for C-array and of 1859+129 for 
D-array, were used
for amplitude and phase calibration. 

The total integration time on CTB 80 was $\sim$ 8 hours in VLA-C
configuration and 3.5 hours in the D-array.
Cygnus A, 8$^{\circ}$ away from the observation center, is a strong source of interference, 
even though it is in the
sidelobes. 
To deal with this problem we need to follow a careful approach because phase errors are
different for sources
in the sidelobes than those in the
mainlobe (due to the nature of noise and how errors propagate in phase and amplitude; see Napier
\& Crane 1982, Perley 1999). 
We thus used the following imaging scheme:
(1) image Cygnus A only; (2) perform a  phase self-calibration using
the model obtained from (1); (3) re-image Cygnus A;
(4) estimate new visibilities's contribution due to Cygnus A as from (3)
and subtract it from the uv-data; (5) image CTB 80 and phase self-calibrate. 
This step was repeated several times and cleaned deeper each time. Finally, (6)
we performed a phase and amplitude self-calibration before
producing the final image. 
The imaging was done with AIPS using the 3-D multi-facet algorithm, with
16 facets, for the C-array data.

The VLA D data, were
initially reduced using standard 2-D imaging techniques,
since the widefield
errors on the quality of the map are expected to be small when the
array is in this configuration. 

The final calibrated C and D maps were combined in the $\textit{u-v}$-domain. 
We used the same two stage imaging as
described in the Sect. 2.1, that is using multi-scale
clean followed by the general 3-D widefield clean. 
The final angular resolution of 
the image at 324.5 MHz is $\sim$ 73$^{\prime\prime}$ $\times$ 
63$^{\prime\prime}$, P.A. = -88$^{\circ}$  and the sensitivity is 14 
mJy beam$^{-1}$.

\subsection{Radio continuum at 618 MHz}

The data at 618 MHz were obtained using the GMRT in two periods on 2002 
May 7 and 8 for a total of 10 hours.
A total of 128 channels were used spread across a bandwidth of 16 MHz.
The observation at this frequency was made in a similar fashion as at 240 MHz 
(see Sect. 2.1). 

The imaging
was done in AIPS++.
We used the wide-field imaging
technique with 9 facets on CTB 80, along with an outlier field on
Cygnus A. 
The large scale structure is faint and is corrupted by the
noise introduced in the deconvolution process. This, we believe, is
due to residual calibration errors with small time scale variations. 
For this image the resulting beamsize is 10$^{\prime\prime}$ $\times$ 
6$^{\prime\prime}$, P.A. = 67$^{\circ}$ and the sensitivity is  
0.6 mJy beam$^{-1}$.

\subsection{Radio continuum at 1380 MHz}

Since the angular size of the SNR CTB 80 exceeds 1$^{\circ}$,  
mosaicing techniques were employed to image the source at 1.4 GHz 
(VLA primary beam $\sim$30$^{\prime}$).
The pointings were separated by 15$^{\prime}$ according to the Nyquist 
sampling criterion.

The observations were carried out in two observing sessions. On January 3, 
1994, 35 pointings were observed with the D configuration of
the VLA in two 50 MHz bands centered at 1385 and 1465 MHz and
on 1995 April 3, 32 pointings were observed with the same configuration of
the VLA, but in two 50 MHz bands centered at 1365 and 1665 MHz.

The data were calibrated using the standard AIPS procedures. The flux density
calibration was based upon  
3C 286  
while the phase calibrator was 1830-210. 
The instrumental polarization was derived using 1830-210, and the
RR--LL offset was calibrated on 3C 286.

The calibrated data from different dates were further processed separately,
using the SDE package (Software Development Environment, Cornwell,
Briggs, \& Holdaway 1996). 
Single antenna data, 
as taken from the Bonn-100 m 1408 MHz Survey
(Reich, Reich, \& F\"urst 1990) were incorporated in the synthesis image.
The single dish data have $T_B(K)/S(Jy)=1.96$ after taking into account an inaccuracy in the
Jy px$^{-1}$ into Jy beam$^{-1}$ conversion in Reich et al.'s (1990) factor. 
This image was employed as the prior image in the maximum entropy based mosaic program
MOSAICM, based on the Cornwell (1988) algorithm. The technique is fully discussed by Holdaway (1999).

Because of the disparity of the second IF in the VLA data (1465 and 1665 MHz) we only
used the low frecuency bands. The images 
observed at 1365 and at 1385 MHz were averaged using
the task COMB in AIPS. The final image  
has an 
angular resolution of $93^{\prime\prime}$ $\times$ $78^{\prime\prime}$, 
P.A. = 72$^{\circ}$ and
an rms noise level of 4 mJy beam$^{-1}$.

\section{Results}

The complex structure of the extended radio remnant CTB 80 
is shown in
Figure 1. 
The top panels show the resulting images of the total intensity
distribution at 240 MHz (Figure 1a) and at 324 MHz (Figure 1b).
The bottom image in Figure 1 shows the SNR as observed near 1380 MHz,
in greyscale and contours (Figure 1c). 
As noted in Sect. 2.3, at 618 MHz
the faint large scale
structures are attenuated due to the lack of short spacings 
and this image is not shown here (see Sect. 5).
Figure 2 shows the new image of CTB 80 at 1380 MHz in a color representation,
useful to display not only the distribution of the bright components, but also 
to illustrate the faint
extensions of the arms revealed for the first time.

From the present data, we have determined the total flux density for 
CTB 80  
S$_{\mathrm 240 MHz}$=99 $\pm$ 15 Jy, S$_{\mathrm 324 MHz}$=85 $\pm$ 9 Jy, and 
S$_{\mathrm 1380 MHz}$=55 $\pm$ 5 Jy.
In all cases, the errors in the flux density measurements take into account the uncertainties 
both in the 
background contribution and in the choice of the integration boundaries
in addition to the inherent rms noises of the images.
These values are based on Perley \& Taylor's (2003) flux density scale. When the observations are made
consistent with the scale of Baars et al. (1977) the flux densities are
S$^{\mathrm B}_{\mathrm 240 MHz}$ = 106 $\pm$ 16 Jy, S$^{\mathrm B}_{\mathrm 324 MHz}$ = 91 $\pm$ 10 Jy
and S$^{\mathrm B}_{\mathrm 1380 MHz}$ = 56 $\pm$ 6 Jy, respectively.
In Figure 3 we present the radio spectrum of CTB 80 from the integrated flux density values 
S$_{\mathrm \nu}$, derived from the new observations (reduced to Baars et al. 1977 flux density
scale for consistence) (black dots) and previous multiwavelength (open symbols) 
estimates taken from the literature
(summarized in Table 2). 
A least-square fit to our data allows to derive a global spectral index of -0.35 $\pm$ 0.16,
in very good agreement with the $\alpha$=-0.35 obtained by Mantovani et al. (1985). A detailed study
of spectral index variations across the SNR and a discussion about a possible spectral break at low
frequencies will be published elsewhere.

One of the major contributions of this work is the high fidelity representation attained in
the 1380 MHz image that has revealed considerable internal structure in the three extended arms,
in spite of
the faintness of the emission.
The N-NE arm 
can be clearly traced to almost R.A. $\sim$ 19$^{\textit{h}}$ 56$^
{\textit{m}}$ to the E, where it bends towards the S and delineates an incomplete circle
(see Figure 2). 
The branching in the emission of this arm around 
R.A. $\sim$ 19$^{\textit{h}}$ 53.5$^{\textit{m}}$, 
decl. between $\simeq$ +33$^{\circ}$ 
05$^{\prime}$ and +33$^{\circ}$ 20$^{\prime}$, is also striking. 
The gap in the emission attains a maximum width of approximately 4$^{\prime}$
near R.A. $\sim$ 19$^{\textit{h}}$ 53$^{\textit{m}}$ 11$^{\textit{s}}$, 
decl. +33$^{\circ}$ 10$^{\prime}$ 
18$^{\prime\prime}$.  
Another striking feature is the 
bright and short NS filament located
along R.A. $\sim$ 19$^{\textit{h}}$
52$^{\textit{m}}$ 56$^{\textit{s}}$, near decl. +33$^{\circ}$ 00$^{\prime}$, that appears 
as a protrusion 
extending to the N from the flat spectrum nebula formed around 
the pulsar. 
The high resolution image of the flat-spectrum ``core'' 
component obtained by Strom (1987) at $\lambda$ 20 cm  
shows an extension of the central nebula (near R.A. $19^h51^m02^s$, 
decl. $32^\circ 
45^\prime 12^{\prime\prime}$, epoch 1950, plate L7 in Strom's paper), in the direction  
of the feature that we detect at large scale.
This correspondence suggests a connection between the two features.
We will refer to this feature  as the ``northern protrusion'' (NP in Figure 4). 
This feature, as well as the bifurcation in the northern arm and 
other morphological characteristics, are evident in the $\lambda$ 92 cm 
image of CTB 80 reported by Strom \& Stappers (2000) based on WSRT observations.
 
The SW wing also exhibits a filamentary structure.  
This portion of the remnant extends to 
declination 
$\sim$ +32$^{\circ}$ 20$^{\prime}$ to the S. 
The presence of a narrow emission arc to the W border (near the center, at about R.A.$\sim$ 
19$^{\textit{h}}$ 52$^{\textit{m}}$, decl.$\sim$ +32$^{\circ}$ 52$^{\prime}$) is
also evident. This filament is separated from the central
nebula by a minimum in emission 
(note Figure 2).
The emission associated with the E wing is also patchy, with brighter spots and voids. 
This is the broadest
of the three extended components, attaining a maximum width of about 11$^{\prime}$.

The three bright point sources that appear overlapping CTB 80 on the N and SE arms
are likely to be extragalactic 
and have been previously identified as
TXS 1952+331, MG3 J195211+3248, and NVSS J195107+323147 (NED database).

\section{Comparison of the 1380 MHz radio emission with infrared and optical features}

\subsection{Radio and Infrared emission}

Based on IRAS HCON images Fesen, Shull, \& Saken  (1988) reported the discovery
of an  IR shell associated with CTB 80. This feature is described as an almost
complete circle,  open to the SW and  centered at 
R.A.(1950)=19$^{\textit{h}}$ 52.9$^{\textit{m}}$, decl.(1950)= +32$^{\circ}$
51$^{\prime}$ ($\pm$ 
2$^{\prime}$) (R.A.(J2000)=19$^{\textit{h}}$ 54.8$^{\textit{m}}$,
decl.(J2000)= +32$^{\circ}$ 59$^{\prime}$), with a diameter of about
64$^\prime$. The structure is revealed by the 
enhanced IR emission ratio 60/100 $\mu$m and appears to the N to
consist of a large arc that matches the N-NE branch of CTB 80.
An increase in  the 60/100 $\mu$m IR color is a useful tracer of 
shock-heated dust. 
Based on HI observations (beam 36$^{\prime}$), Koo 
et al. (1990) report the detection of an expanding HI shell with a center, 
shape and neutral gas mass that matches the properties of the IR shell.
Later, on the basis of HI observations carried out with higher 
angular resolution (beam of $\sim$
1$^{\prime}$ and 3$^{\prime}$) Koo et al. (1993) conclude that the 
interstellar medium around CTB 80 consists of cold neutral
clumps immersed in a warm neutral medium and suggest that the SNR has 
apparently encountered a cavity to the SE (that coincides
with enhanced IR color). 

Here we present a comparison of the new $\lambda$20 cm image with the 
IR color distribution. Figure 4
displays an overlay of the radio contours at 1380 MHz 
with the 60/100 $\mu$m ratio image constructed from the
$\textit{IRAS Sky Survey Atlas}$. We find that the 60/100 IR color is 
enhanced  behind the radio shock
front in all the extension of the N-NE arm of the SNR (as shown in Figures
1 and 2), in agreement with the conclusions of Fesen et al. (1988) and
Koo et al. (1993). The quality of the new $\lambda$20 cm radio image 
allows us to confirm this association and to establish the IR/radio 
agreement up to $\sim 19^{\textit{h}} 55^{\textit{m}} 50 ^{\textit{s}}$
to the E (Koo et al. 1993 had noticed the agreement only up to
positions near 19$^{\textit{h}} 54.5^{\textit{m}}$ based on the comparison with
a $\lambda$49 cm image, see for example Figure
9 in their paper). Moreover, in the place where
the radio emission associated with the N arm of CTB 80 curves to the S 
in an arc at the NE extreme,
delineating a faint, open  circle (centered near 
R.A.$\simeq$ 19$^{\textit{h}}$ 55$^{\textit{m}}$, 
decl.$\simeq$ +33$^{\circ}$ 15$^{\prime}$; see for example in the color
image in Figure 2), the IR 
enhancement perfectly matches this peculiar curved
morphology. This last IR feature, 
weaker than the rest of the shell, 
has a color
corrected ratio I$_{60\mu m}$/I$_{100\mu m}$ $\simeq$ 0.30, thus implying
a dust temperature of the order of
26 K, in agreement with the T$_{d}$=27.5$\pm$2.0 K derived by  Fesen et al.
(1988) for the large IR shell, and  compatible with shock heated dust 
temperatures as observed in other SNRs (Arendt et al. 1992). We conclude
that this singular termination observed in the N arm is real, and has the
same nature as the rest. 
In addition to the general IR/radio correspondence in this part of the remnant,
two aspects of the inner
structure of the N arm of CTB 80 are striking: (1) the radio  
``northern protrusion'' (NP in Figure 4),
has an infrared counterpart that mimics the appearance
and extension of the radio enhancement; (2) the two branches observed in the
radio emission near R.A. 19$^{\textit{h}}$ 53.5$^{\textit{m}}$,
decl. +33$^{\circ}$ 11$^{\prime}$ correspond with local maxima in the 
IR emission.

In conclusion, the present accurate IR/radio comparison provides
additional support to Fesen et al's. (1988) and Shull et al.'s (1989)
suggestions that the pulsar's relativistic electrons may  rejuvenate
an old shell, giving rise to the synchrotron emission where the magnetic
field lines have been compressed behind radiative shocks. In addition,
the morphology and brightness of the `northern protrusion'' suggest that
this structure might be tracing
the link between the pulsar and the extended N 
radio continuum arm.  

In contrast to the excellent agreement found in the N branch,
little correlation between IR and radio continuum features 
 can be shown from the morphological point of view in the rest of the
proposed IR shell around CTB 80. 

An interesting IR feature is the roughly elliptical ring of higher IR 60/100
ratio located to the south of CTB 80,  centered near 
19$^{\textit{h}}$ 55$^{\textit{m}}$, +32$^{\circ}$ 30$^{\prime}$.  
In displays of the 1380 MHz radio emission at the level of   4-sigma 
($\sim$ 12 mJy/beam), a faint radio 
structure (with maxima of $\sim$ 21 mJy/beam) is detected at the
same location as this IR ring, with identical shape and diameter. 
No similar feature
has been, however, detected at 240 and 324 MHz above the respective
noise levels, suggesting that this feature may be thermal.
As noticed by
Mavromatakis et al. (2001) this IR feature is associated with optical filaments
(see Figure 5 in the next section). 

\subsection{Radio and Optical emission}

Several optical studies (narrow-band imagery and spectroscopy) have shown that 
the remnant consists of outlying diffuse and filamentary emission
surrounding a central ring of filaments, coincident with the central radio
emission peak (Blair et al. 1984 and references therein, Fesen \& Gull 1985). 
Recently, Mavromatakis et al. (2001) have presented new optical images from deep CCD 
exposures in the lines
H$_{\alpha}$+[NII] $\lambda$6555 $\AA$, [SII] $\lambda$6708 $\AA$, [OII] 
$\lambda$3727 $\AA$, and [OIII]
$\lambda$5005 $\AA$
covering a 2$^{\circ}$ $\times$  2$^{\circ}$ area. 
These large scale
images 
reveal new filamentary and diffuse structures to the S, SE, and N of CTB 80.
Here we use these images to analyze the large-scale emission, far from the pulsar nebula.

Figure 5 shows an overlay of the radio continuum emission at 1380 MHz with 
the [SII] optical emission.
Two enlargements of interesting areas (of size $\sim$ 30$^{\prime}$) are included to 
facilitate 
the detailed comparison.
The radio and optical features along the SW wing are well correlated, especially in
the filament labeled I in Mavromatakis et al. (2001) (see close-up image to the right), 
where the optical filaments
closely match the radio emission. In addition 
the radio/IR feature ``NP'' has an optical counterpart (see the top close-up image).
The optical emission also reproduces the branching noticed in both radio and 
IR in the N-NE arm of CTB 80. The
[SII] emission accompanies the radio synchrotron emission all along this arm. 
It is interesting to note that at the eastern extreme of this arm, near R.A. $\simeq$ 19$^{\textit{h}}$
55$^{\textit{m}}$ 30$^{\textit{s}}$, decl. $\simeq$ 33$^{\circ}$ 20$^{\prime}$, short, bright optical 
filaments 
are observed in coincidence with a flattening in the radio contours and enhanced IR 60/100
ratio. It is possible that the expanding shock has encountered denser material at this location,
of the kind of dense cold clumps reported by Koo et al. (1993) in the SNR shell.
A dense cloud at this location may have modified the shock expansion, originating the singular 
``curl'' observed at the termination of this arm.
The bright larger optical filaments observed 
farther east (R.A. $\sim$ 19$^{\textit{h}}$ 57$^{\textit{m}}$, 
decl. $\sim$ +33$^{\circ}$ 15$^{\prime}$)
correspond to the HII region LBN 156 (Lynds 1965). 

While the SW filaments are conspicuous in all the optical lines, 
the N-NE arc is only prominent in [SII] line.
[SII] is known to be a tracer of hot, shocked gas in regions where the expanding SNR 
shell encounters
and overtakes enhancements in the surrounding gas (Fesen et al. 1997).
This fact is compatible with the presence of the IR and HI shell in this portion of the SNR.
In all cases, the complex of optical filaments are detected behind the shock front as 
delineated by the radio synchrotron emission. 
According to Draine \& McKee (1993), the presence of shocked interstellar gas 
(delineated by 
H$_{\alpha}$ and [SII] emission) indicates the presence of radiative shocks. Regions behind such
shocks undergo strong compression, resulting in both magnetic fields amplification and 
enhancements
in the density of relativistic particles. The excellent morphological correspondence observed
in CTB 80 confirms
this scenario.

The filaments  III and IV from the images by Mavromatakis et al. (2001)
lie to the S of CTB 80 and are probably unrelated. The filaments called V and VI perfectly match
the E, S, and W sides of the IR hollow feature shown in Figure 4 to be located S of CTB 80, around
19$^{\textit{h}}$ 55$^{\textit{m}}$,
+32$^{\circ}$ 30$^{\prime}$. Mavromatakis et al. (2001) suggest that these filaments are also
related to CTB 80. On the basis of the new sensitive radio observations, we conclude that their
association with CTB 80 is unlikely. Further deep multispectral studies of this region are needed
to confirm the nature of the observed emissions at optical and IR wavelengths.

\section
{The Pulsar Wind Nebula and its X-ray counterpart}

Angerhofer et al. (1981) have characterized the central emission in CTB 80 as composed of a
compact $\sim$ 45$^{\prime\prime}$ ``core'' nebula immersed in a 10$^{\prime}$ $\times$ 6$^{\prime}$
``plateau'' emission.
Since both components
have flat spectra, compatible with a nebula powered by the pulsar wind
(Castelletti et al. 2003, in preparation), we use in what follows
the term PWN to describe both, the ``plateau'' and ``core'' components.

The highest resolution radio image for the plateau nebula that contains the 
pulsar PSR B1951+32 at the W end was obtained at 618 MHz 
(beam 10$^{\prime\prime}$ $\times$ 6$^{\prime\prime}$, PA= 67$^{\circ}$.5).
This image improves by a factor of 10 the angular resolution of Angerhofer et al.'s (1981) image
of this region of the SNR. Figure 6 displays in 
greyscale and white contours the synchrotron emission at 618 MHz.  
The \it ROSAT \rm
PSPC X-ray data (1 to 2.4 keV) are shown in black contours (Safi-Harb et al. 1995). 
The X-ray data are smoothed by a 72$^{\prime \prime}$ Gaussian following
Safi-Harb et al. (1995).
In spite of the presence of low level interferometric artifacts (striations at an angular scale
of a few arcsec), this image reveals that the radio emission in the ``plateau'' nebula is structured and 
extends $\sim$ 8$^{\prime}$ $\times$ 4$^{\prime}$ (5 pc $\times$ 2 pc at the distance of 2 kpc). 
There are two  prominent features in Figure 6 with intensities about five times
brighter than the surrounding levels: one running along the nebula, approximately
aligned in the E-W direction
and the other located to the SW of the pulsar. The E-W feature appears to
trail from the pulsar and has a morphology suggestive of a helix. 
This source has a size of
about 6$^{\prime}$ ($\sim$ 3.5 pc). 
The narrow 
feature at the SW of the nebula has an extension of $\sim$ 3$^{\prime}$.7  ($\sim$ 2 pc). 
An absolute maximum of 110 mJy beam$^{-1}$ is observed
at the pulsar position.
From the comparison of the ``plateau'' radio nebula with X-ray emission, two 
characteristics are apparent:
(1) the X-ray nebulae (both the core and the
extended plateau) have comparable sizes 
to the radio features, and both
are asymmetric with extensions in the E-W direction; (2) although
the radio and X-ray are similar in size and general appearance,
deviations are observed in the E extreme: the radio nebula
shows a bend to the S, while the X-rays emission extends to the N.

A blow-up of our image at 618 MHz of the  
``core'' radio nebula is shown in 
the lower panel of Figure 6. In this enlargement, we display in black contours the 618 MHz emission,
and in greys the central nebula as observed at 1.4 GHz by Strom (1987) with 
$\sim$1$^{\prime\prime}$ resolution, in order to facilitate the comparison 
of the information at the different angular scales.
The position of the pulsar is marked by the plus sign,
and the arrow shows the direction in which the pulsar is moving 
according to Migliazzo et al. (2002).
The ``core'' nebula appears slightly elongated in the E-W direction
with a size of about 1$^{\prime}$.2 
($\sim$ 0.7 pc). 
In the close-up image included in Figure 6,
a distortion towards the western edge 
with a position angle of
259$^{\circ}$ (N through E) is also observed. 
This position agrees within errors with the estimate
by Migliazzo et al. (2002) for the pulsar motion  
PA=(252$^{\circ}$ $\pm$ 7$^{\circ}$). 
In general X-rays and radio radiation are in good agreement; however 
there is
no radio counterpart for an extended ``conelike'' hard X-ray
emission feature located SE of the pulsar, reported by Safi-Harb et al. 
(1995).

\section{Energetics in CTB 80 and in the PWN}

Based on the observed properties, we can estimate the energy content of CTB 80. 
The synchrotron energy consists of the   
kinetic energy of the relativistic particles and the energy stored in the electric and 
magnetic field (i.e. Poynting flux). 
Assuming equipartition between particles and magnetic field we estimated through the relation 
derived by Moffet (1975) a minimun energy 
content of relativistic 
electrons for CTB 80 in the range
$\simeq$ 5.5 $\times$ 10$^{48}$ to 5 $\times$ 10$^{49}$ ergs, for a ratio between the energy
in relativistic electrons and energetic baryons assumed to be 1 and 50, respectively. 
These values, low as compared with the canonical 10$^{51}$ ergs for SN, suggest that the 
energy required to accelerate
electrons represents only a small fraction of the total energy. Most of the
energy must have been dissipated in the interstellar medium as kinetic energy.

For the central components (core and plateau), based on the size of the radio 
nebula (assuming it is formed by the central core plus the
component
trailing to the E)
and the pulsar velocity of 240 km s$^{-1}$ derived
by Migliazzo et al. (2002), we can estimate 
an age of
18200 yr, for the PWN. 
Based on the average Lorentz factor of the electrons, Safi-Harb et al. (1995) conclude that the
lifetime of the X-ray emitting electrons in the plateau nebula is 
$\sim$ 20000 yr, in good agreement with the current radio results.
By assuming equipartition, we find
that the energy content in relativistic particles in the PWN (plateau plus core) 
is in the range 1.3 $\times$ 10$^{47}$ to 1.2 $\times$ 10$^{48}$ ergs (for an energy ratio between 
electrons and baryons of 1 and 50, respectively). 
The spin-down energy loss rate for PSR B1951+32 has been estimated as 
\textit{\.E} = 3.7 $\times$ 10$^{36}$
ergs sec$^{-1}$ (Kulkarni et al. 1988). Thus the energy injected by the pulsar during a total lifetime
of $\sim$ 64 kyr (Migliazzo et al. 2002) is expected to be \textit{E} $\sim$ 7 $\times$ 10$^{48}$
ergs, in reasonable agreement with the energy content observed in relativistic electrons in the PWN
when equipartition
is assumed.

We have estimated the magnetic field of the plateau nebula 
$B(\mu$G)=1040($\nu_{br}$/10$^{12}$Hz)$^{-1/3}$ ($t_{neb}$/1000 yr)$^{-2/3}$
(Pacholczyk 1970).
To estimate the cutoff frequency $\nu_{br}$, we follow Chevalier's (2003) analysis for 
3C 58 and MSH 15-52.
Chevalier (2003) suggests that the fact that the X-ray nebula extent is comparable to that of the 
radio nebula
indicates that $\nu_{br}$ is not much lower than X-ray energies, assuming that particles 
originate
close to the pulsar and then disperse in the nebula. The case for the extended nebula 
associated with PSR B1951+32
is analogous. 
We thus adopt a break frequency $\nu_{br}$ $\approx$ 2.4 $\times$ 10$^{16}$ Hz.
Based on this frequency and the estimated age for the nebula \textit{t$_{neb}$} $\simeq$ 18200 yr, 
the nebular
magnetic field is $\simeq$ 5.2 $\mu$G.
This field is in reasonable agreement with the $\sim$ 3.4 $\mu$G derived by Safi-Harb et al. 
(1995) based on
X-ray data, and comparable to the 3C 58 and MSH 15-52 magnetic fields (16 and 8 $\mu$G, respectively).

On the other hand, the total pressure of the plateau component of the PWN can be estimated as 
$P_{T} \sim $ (2/3)
($E_{min}/V_{0}$), where $V_{0}$ is the source volume. 
We obtain a total pressure of 
$ \simeq 5.2 \times 10^{-11} $ to $ 4.8 \times 10^{-10} \mathrm{dyn\, cm^{-2}}$,
which easily satisfies the condition for confinement of the 
wind 
($\rho V^{2}_{PSR}$ $>$ $P_{T}$) for an  ambient
density of $\simeq$ 0.5 H atoms cm$^{-3}$. For the
core region, the total pressure is in the range $2.1 \times 10^{-10} \mathrm{dyn\, 
cm^{-2}}$ to 
$ 1.9  \times 10^{-9} \mathrm{dyn \, cm^{-2}}$. 

\section{Summary}

We have presented new images of the SNR CTB 80 at 240 MHz, 324 MHz, 618 MHz, and 
1380 MHz, using  the GMRT and VLA.
The radio images obtained at all the different frequencies display 
the same underlying morphology.
The new images reveal the faint wings of CTB 80 in their full extent. 
The northern arm exhibits a peculiar
bifurcation in the emission, with a low intensity gap between the two edges. 
Also a bright feature, called 
the ``northern protrusion'' appears as an extension of the pulsar wind nebula to the north,
suggesting that this may be the link between the pulsar energy and the radiation in the extended N arm of
CTB 80.
The SW arm also exhibits a filamentary structure. 
The central E-W component is the broadest and is patchy in
appearance.
From the comparison of the high resolution and sensitivity radio data at 1380 MHz with the IR shell 
proposed by
Fesen et al. (1988) to be associated with CTB 80, we conclude that there is an excellent agreement
between radio continuum emission and IR color all along the N arm of the SNR.  
Moreover,
with the sensitivity attained at 1380 MHz the N arm of CTB 80 is shown to have a peculiar curved
termination at its eastern extreme, in concordance with a similar feature in the IR.
Little IR/radio correspondence is, however, observed associated with the E and SW arms of CTB 80.
Optical filaments prominent in the [SII] line are observed to accurately match the radio emission
along the W and N edges of CTB 80. In all cases, the optical emission is detected behind the shock
front as delineated by the radio emission.

The pulsar wind nebula (the ``core'' plus ``plateau'') was investigated with arcsec resolution at 
618 MHz. 
The new radio image shows with 
unprecedented detail the presence of a helicoidal bright trail behind the pulsar ``core'' nebula
and a distortion in the ``core'' nebula in the direction of motion of the pulsar
PSR B1951+32.
The correspondence between the radio synchrotron nebula and the 1-2.4 keV X-rays
associated with the PWN is also striking.
However  
the images in the X-ray and radio diverge in the plateau at about 8$^\prime$ E from the pulsar.
In effect, while the radio emission at all the observed
frequencies is clearly aligned in the E-W direction with a slight bend  
to the S, the X-ray
nebula covers approximately the same extension but curves to the N.

From the present data we have estimated the total energy carried by relativistic 
electrons in CTB 80 to lie in the range 
5.5 10$^{48}$ to 5 $\times$ 10$^{49}$ ergs; while for the PWN the energy content is
between
1.3 $\times$ 10$^{47}$ and 1.2 $\times$ 10$^{48}$ ergs, depending on the particle composition. 
The total pressure in the 
nebula has been estimated $\simeq$ $5.2 \times 10^{-11}$ to $4.8 \times 
10^{-10} \mathrm{dyn \, cm^{-2}}$.
A magnetic field of about 5.2 $\mu$G has also been estimated for the PWN.

A subsequent paper (Castelletti et al. 2003 in preparation) will deal with 
local spectral index variations in the entire source.
The study of spectral distribution may help to understand the role of the pulsar in influencing 
the shape and energetics
of this peculiar SNR.

\acknowledgements

We thank the staff of the GMRT that made these observations possible; in particular 
we would like to thank
N. Kantharia 
and S. Roy.  We are grateful to F. Mavromatakis for providing the optical 
images, to
S. Safi-Harb for providing the  X-ray image, and to R. Strom for the 49 cm data. We wish to thank the
referee whose constructive criticism has helped to make this a better paper. G. C. and P. F. V.  
acknowledge the support
and hospitality from NRAO during their stay at the VLA. This research was partially funded through 
CONICET
(Argentina) grant 4203/96 and the UBACYT grant A013 (Argentina). PFV is supported by CONACYT 
grant 36572-E and DGAPA-UNAM grant IN 112602.
This research has made use of the NASA's ADS Bibliographic Services.

\begin{figure}
\plotone{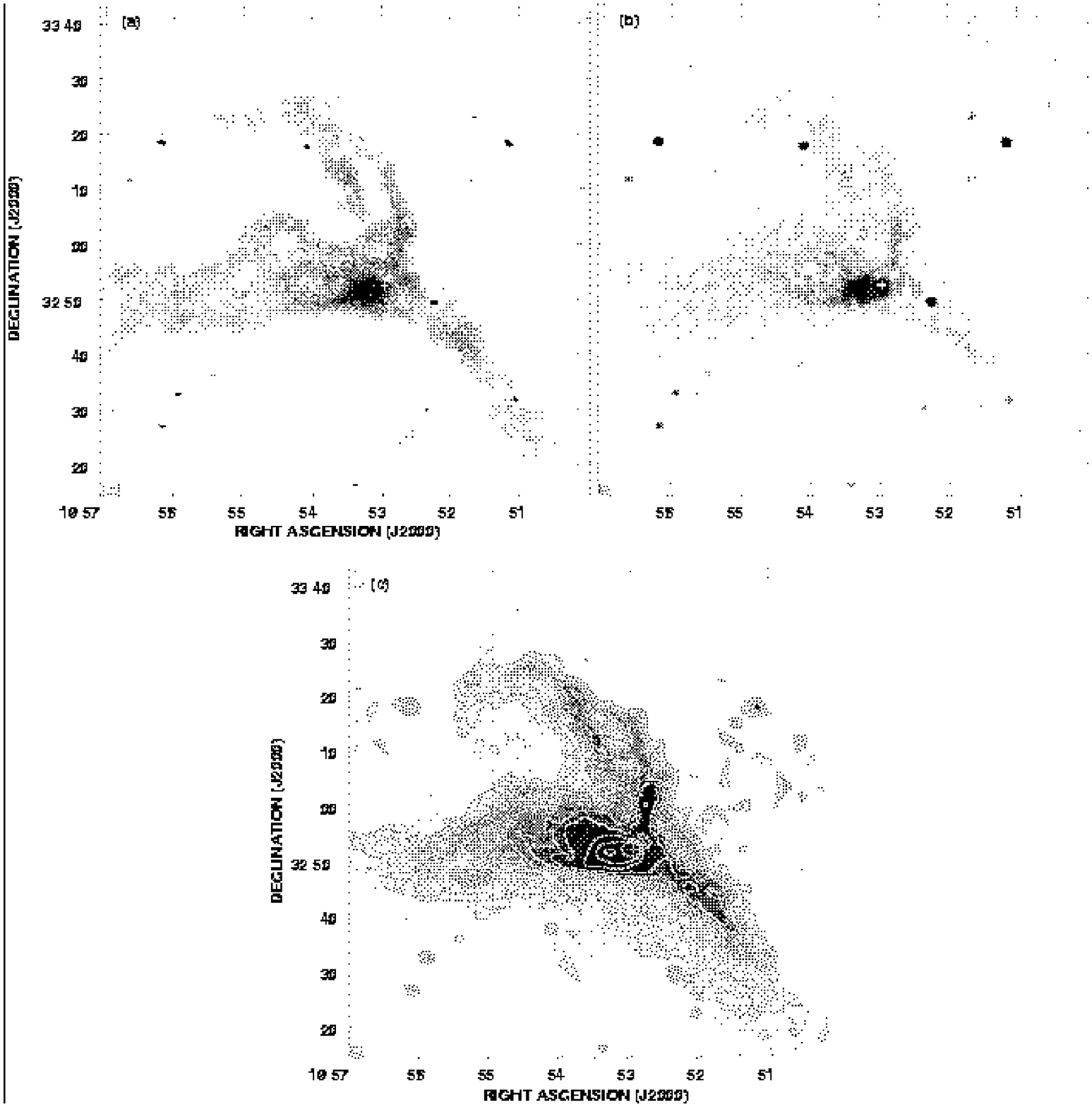}
\caption{Extended emission of SNR CTB 80 at $\textit{(a)}$ 240 MHz, $\textit{(b)}$ 324 MHz, and 
                 $\textit{(c)}$ 1380 MHz.
                 Fig. $\textit{1a}$ displays the radio emission
                 at 240 MHz. The angular resolution 
                 is 36$^{\prime\prime}$ $\times$ 23$^{\prime\prime}$,
                 and the rms noise = 4.6 mJy beam$^{-1}$.
                 The grey-scale varies between 7 and 45 mJy beam$^{-1}$.
                 Fig. $\textit{1b}$ shows the 324.5 MHz image.
                 The beamsize is 73$^{\prime\prime}$ $\times$
                 63$^{\prime\prime}$, P.A. = -88$^{\circ}$, rms noise =  14 mJy beam$^{-1}$.
                 The grey-scale ranges from 25 to 400 mJy beam$^{-1}$.
                 Fig. $\textit{1c}$ shows the emission at 1380 MHz.
                 The angular resolution is 93$^{\prime\prime}$ $\times$
                 78$^{\prime\prime}$, P.A.= 72$^{\circ}$, and the rms
                 noise level is 4 mJy beam$^{-1}$.    
                 The grey-scale
                 ranges from 20 to 120 mJy beam$^{-1}$
                 and the contour levels are
                 25, 30, 45, 55, 65, 80, 100, 150, and 250 mJy beam$^{-1}$.
                 In all cases, the position of the pulsar PSR B1951+32 is indicated by a plus (+)
                 sign.
                 The beamsizes are included in the bottom left-hand corner
                     of each image.}
\end{figure}

\begin{figure}
\plotone{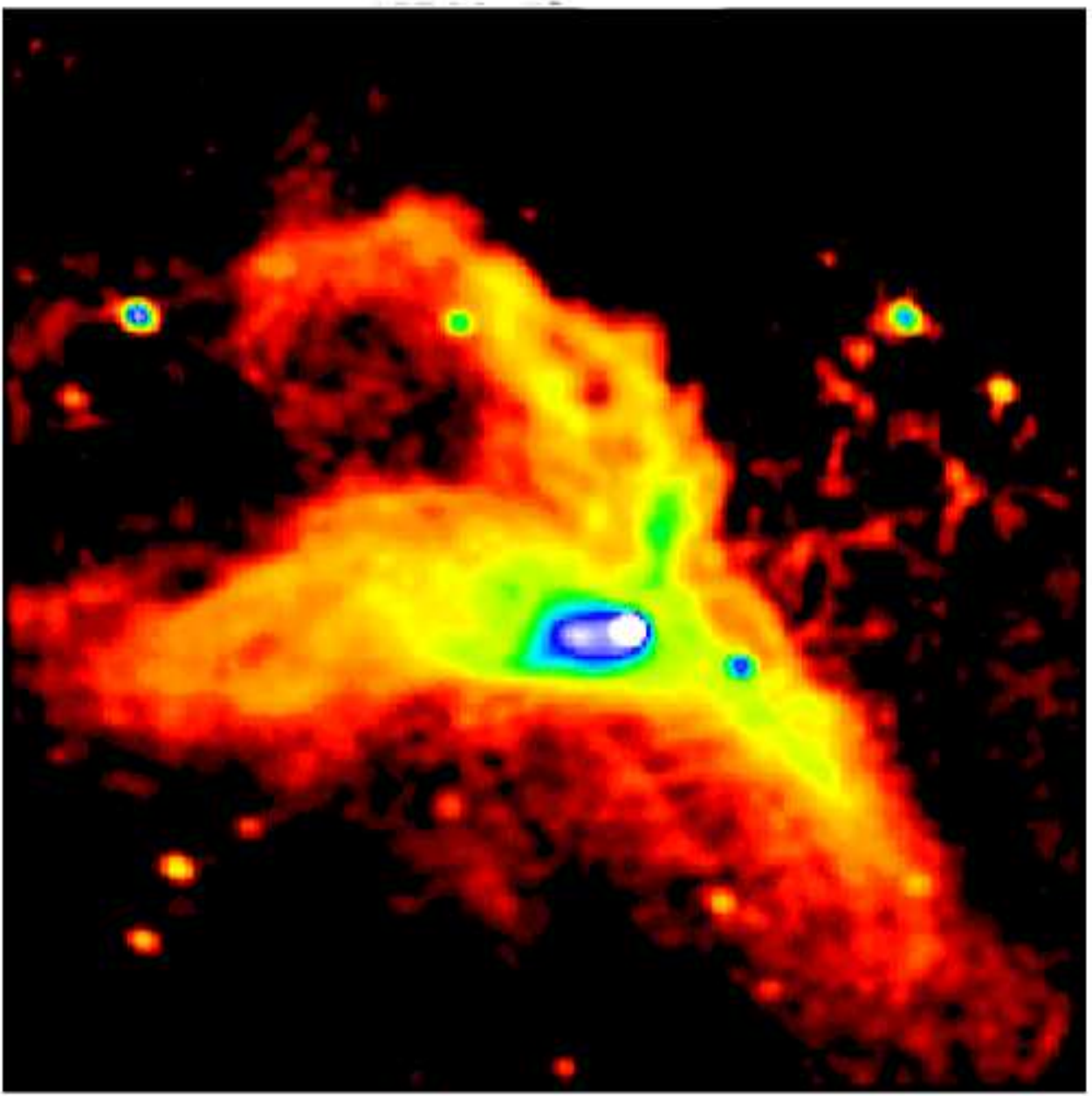}
\caption{Pseudocolor representation of the SNR CTB 80 at 1380 MHz.
             This image was obtained from a
             combination of 32 different pointings observed with the VLA in the D configuration.
             Single-dish data at 1408 MHz, from the Bonn-100 m observations, have been added
             to the interferometric image. The angular resolution is 93$^{\prime\prime}$ $\times$
             78$^{\prime\prime}$, P.A.= 72$^{\circ}$. The rms noise level is 4 mJy beam$^{-1}$.
             The brightness range covered by the color scale ranges from 20 to 600 
             mJy beam$^{-1}$.}
\end{figure}

\begin{figure}
\plotone{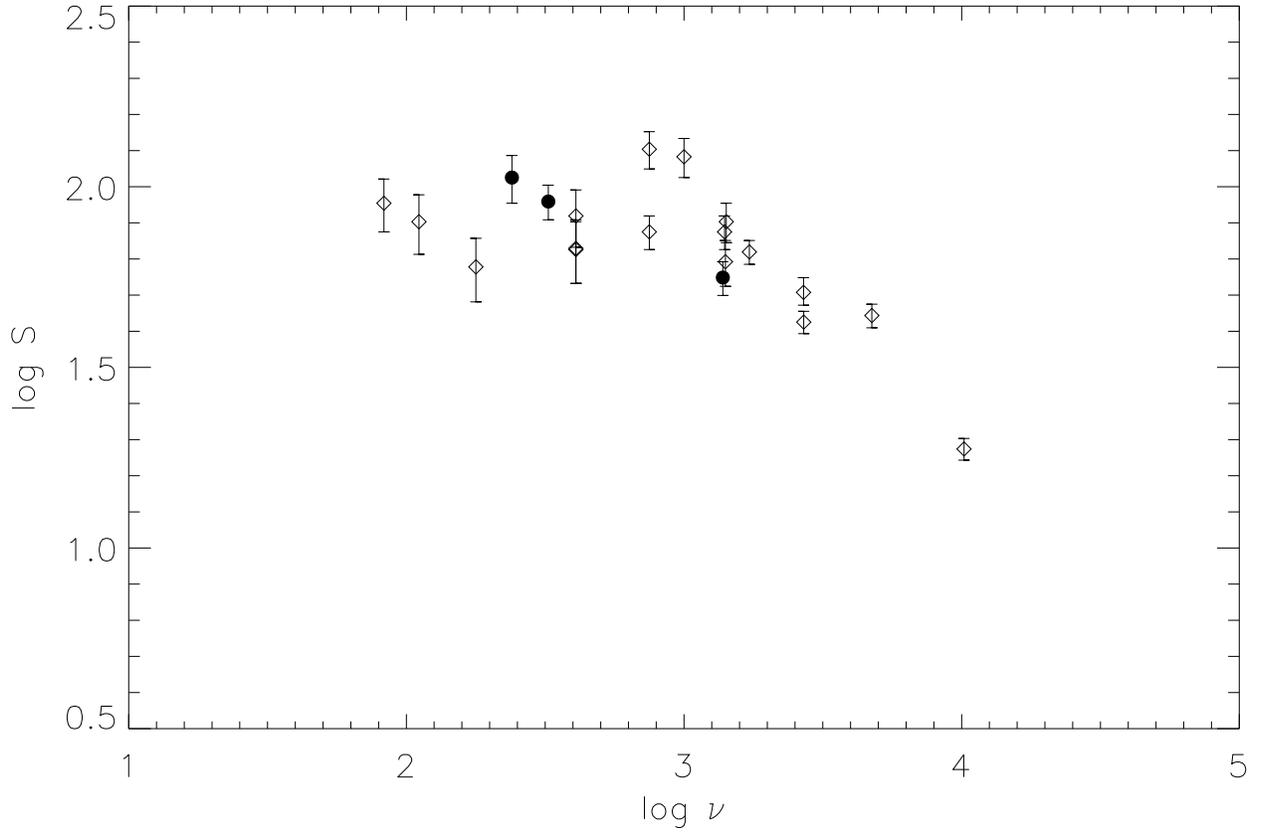}
\caption{Radio spectrum of CTB 80. Open symbols represent data taken from the literature,
          listed in Table 2,
         and the black dots correspond to the present data at 240 MHz, 324 MHz and 1380 MHz.}
\end{figure}

\begin{figure}
\plotone{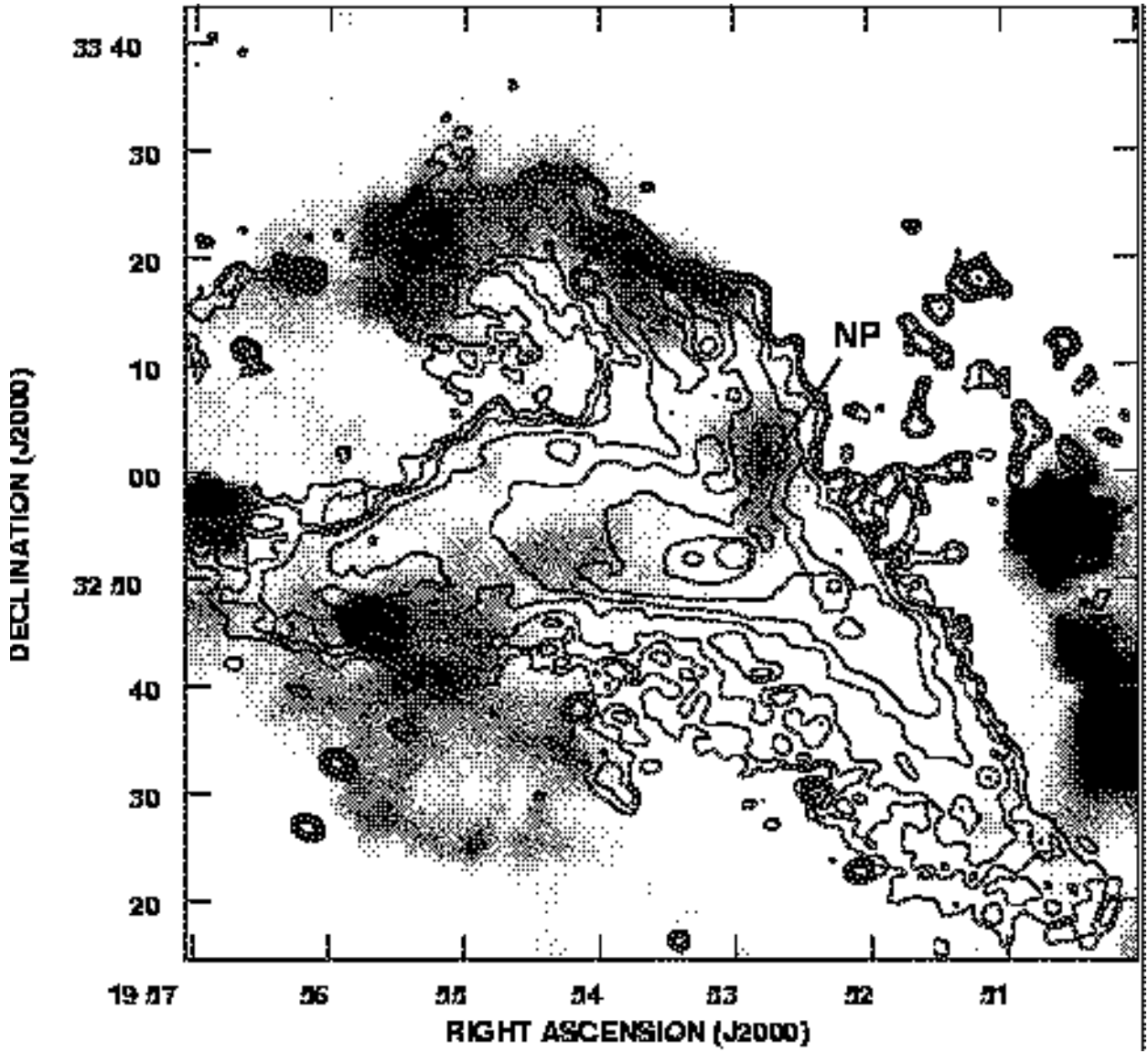}
\caption{Infrared emission in the CTB 80 complex region. The gresyscale distribution
         corresponds to the 60 $\mu$m/100 $\mu$m infrared ratio. The plotted contours 
         at 1380 MHz are at 21, 25, 30, 45, 65, 100, 230, 370 mJy beam$^{-1}$.
         The greyscale in the IR ratio varies between 216 and 235 milli-ratio.
         The location of the northern protrusion (NP)
         feature is indicated.} 
                
\end{figure}

\begin{figure}
\plotone{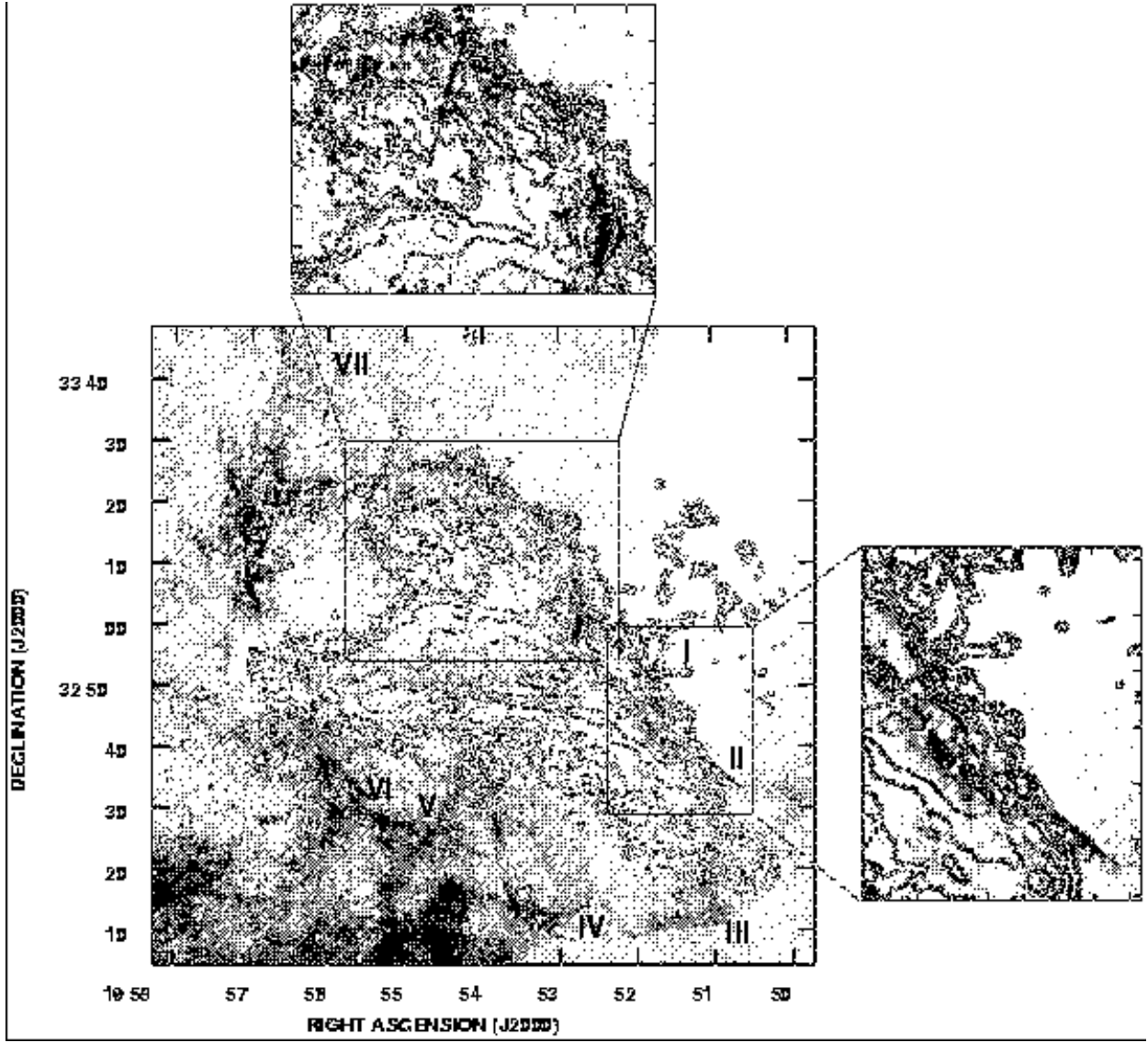}
\caption{Comparison of the optical emission in the [SII] line (greyscale representation)
            as taken from Mavromatakis et al. (2001) with the
               continuum emission from the VLA at 1380 MHz. The radio contours are at
               21, 25, 30, 45, 65, 100, 230, 370 mJy beam$^{-1}$.
               The roman numbers, in the central image, correspond
               to the designation of areas by Mavromatakis et al.
               (2001). Close ups of two interesting areas in the [SII] image are displayed 
                in the top and right panels with
                some radio contours
                overlapping.}
\end{figure}

\begin{figure}
\plotone{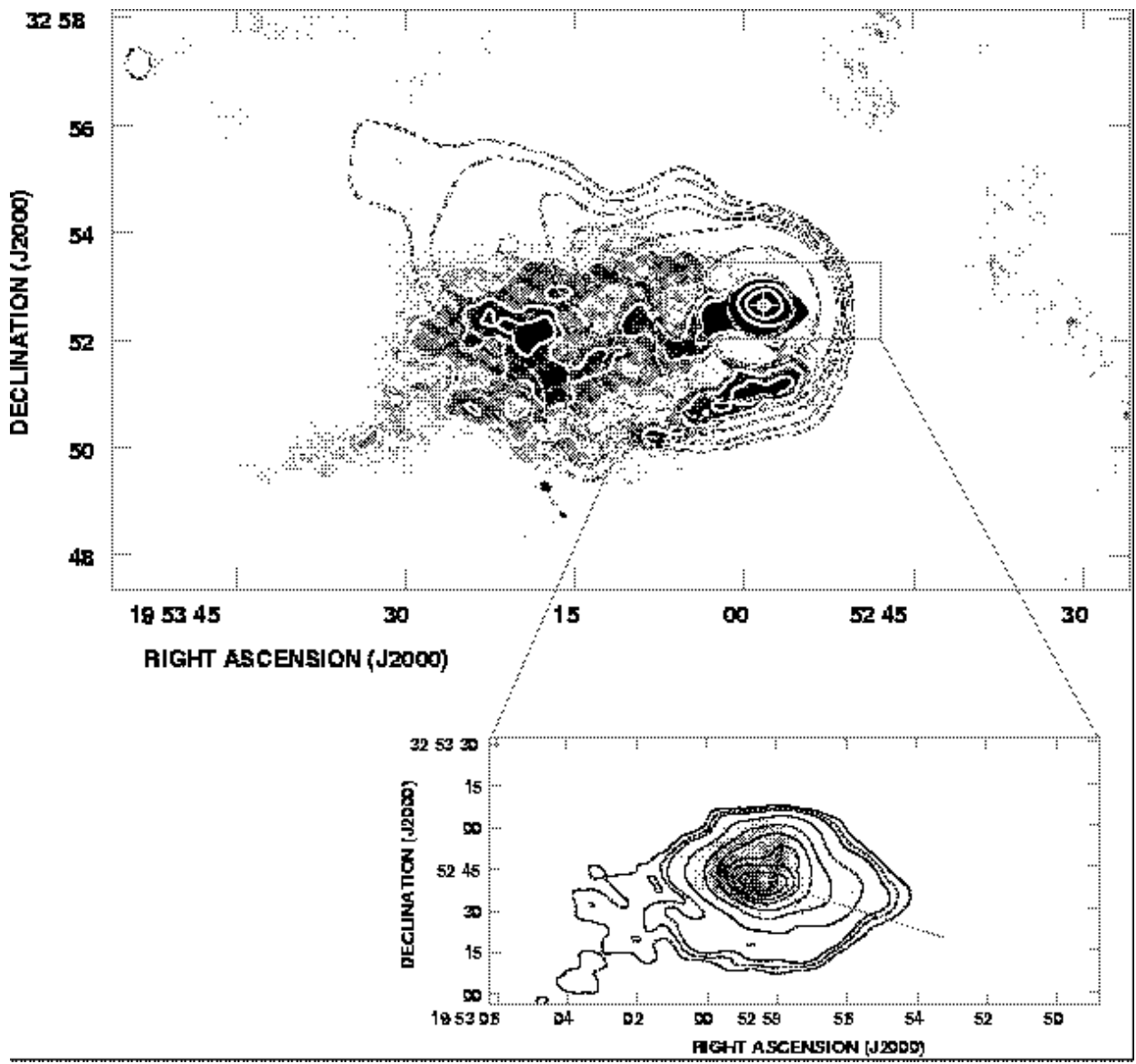}
\caption{Overlay of the central component of CTB 80 as imaged
            at 618 MHz (grey and white cotours) with the \it ROSAT \rm PSPC X-ray data
            in the photon energy band
            1-2.4 keV (black contours). The radio image is smoothed to a 
            resolution of
            12$^{\prime\prime}$ $\times$ 8$^{\prime\prime}$, and the rms
            noise level is 0.6 mJy beam$^{-1}$. The greyscale runs from 1 to 5 mJy beam$^{-1}$
            and the white contours
            are at 2, 3, 4, 6, 8, 30, 70, 130 mJy beam$^{-1}$.
            The lower panel is a close-up image of the core region in CTB 80. The image
            at 1382.5 MHz (greyscale from 0.08 to 2 mJy beam$^{-1}$) as taken from
            Strom (1987) has a resolution of 1$^{\prime\prime}$,
            P.A.=89$^{\circ}$.53. The
            618 MHz image is shown in contours traced at
            6, 8, 10, 20, 40, 60, 80, 90, 100, 120 mJy beam$^{-1}$.
            The plus (+) sign shows the position of the pulsar PSR B1951+32 and the arrow
            indicates the direction of its proper motion as derived by Migliazzo et al. (2002).}

\end{figure}

\clearpage

\begin{deluxetable}{lll}
\renewcommand{\arraystretch}{.73}
\tablenum{1}
\tablewidth{0pc}
\tablecaption{Observational parameters}
\tablehead{
\multicolumn{3}{c}{GMRT Observations}}
\tablecolumns{3}
\startdata
Central frequency & 240 MHz  & 618 MHz \\
Field center (J2000)& 19 53 21, +32 55 42 & 19 53 21, +32 55 42\\
Observing dates & May 6, 7 2002 &  May 7, 8 2002 \\
Min. and max. baselines & 52 m - 26 km  & 52 m - 26 km \\
Bandwidth  & 8  MHz & 16 MHz\\
Total observing time & 10 hours & 10 hours \\
Main Calibrators and assumed  & 3C 48 (50 Jy) & 3C 48 (29 Jy) \\
\hspace{1cm}flux densities \tablenotemark{a} & 3C 286 (28 Jy) & 3C 286 (21 Jy) \\
FWHM of primary beam & 150$^{\prime}$ & 54$^{\prime}$\\
Synthesized beam & 26$^{\prime\prime}$ (EW) $\times$ 17$^{\prime\prime}$ (NS) &
                 10$^{\prime\prime}$  (EW)  $\times$ 6$^{\prime\prime}$ (NS)\\
Position angle (E to N) & 76$^{\circ}$.92 & 67$^{\circ}$.46 \\
Noise level &  4.6 mJy/beam& 0.6 mJy/beam\\
\cutinhead{VLA Observations}
Central Frequency & 324  MHz & 1380 MHz \\
Configurations & C and D &
D and single dish \tablenotemark{b}\\
Field center (J2000) & 19 52 58, +32 55 41 & 19 55 42, +32 32 33\\
Observing dates & March 21 2000, Aug. 14 2000& Jan. 3 1994, April 3 1995\\
Min. and max. baselines &35 m - 3.3 km & 35 m - 1.03 km\\
Bandwidth  & 3 MHz & 50 MHz \\
Total observing time & 11.5 hours & 7.5 hours \\
Main Calibrators and assumed & 3C 286 (25 Jy) &  3C 286 (14.55 Jy)   \\
\hspace{1cm}flux densities \tablenotemark{a} &  3C 48 (42.8 Jy) &  \\
FWHM of primary beam & 139$^{\prime}$ & 32$^{\prime}$ \\
Synthesized beam & 73$^{\prime\prime}$  $\times$ 63$^{\prime\prime}$  &
93$^{\prime\prime}$  $\times$ 78$^{\prime\prime}$   \\
Position angle (E to N) & -88$^{\circ}$ & 72$^{\circ}$ \\
Noise level  &  14 mJy/beam & 4 mJy/beam \\
\enddata
\tablenotetext{a}{VLA Calibrator Manual 2003,
(http://www.aoc.nrao.edu/$\sim$gtaylor/calib.html).}
\tablenotetext{b}{Data taken from the Bonn-100 m 1408 MHz Survey
(Reich, Reich, \& F\"urst 1990).}
\end{deluxetable}

\clearpage
\begin{deluxetable}{rrr}
\renewcommand{\arraystretch}{.73}
\tablenum{2}
\tablecolumns{3}
\tablewidth{0pc}
\tablecaption{Integrated flux density estimates for CTB 80}
\tablehead{
\colhead{Frequency} &\colhead{Integrated Flux} & \colhead{References}\\
\colhead{(MHz)}&\colhead{(Jy)}&\colhead{}}
\startdata
83 & 90 $\pm$ 15 & Kovalenko, Pynzar, \& Udal'tsov (1994)  \\
111 & 80 $\pm$ 15 &  Kovalenko, Pynzar, \& Udal'tsov (1994) \\
178 & 60 $\pm$ 12 & Bennett (1963) \\
240 & 106 $\pm$ 16& Present work \tablenotemark{a} \\
324 & 91 $\pm$ 10 & Present work \tablenotemark{a} \\
408 & 67 $\pm$ 13 & Felli et al. (1977) \\
408 & 67.5 $\pm$ 10.5 & Mantovani et al. (1985) \\
408 & 83 $\pm$ 15     & Haslam et al. (1982) \\
750 & 127 $\pm$ 15    & Velusamy, Kundu, \& Becker (1976) \\
750 & 75 $\pm$ 8 & Pauliny-Toth, Wade, \& Heeschen (1966) \\
1000 & 121 $\pm$ 15 & Velusamy, Kundu, \& Becker (1976)\\
1380 & 56 $\pm$ 6 &  Present work \tablenotemark{a} \\
1400 & 75 $\pm$ 8 & Pauliny-Toth, Wade, \& Heeschen (1966) \\
1410 & 62 $\pm$  9 & Mantovani et aal. (1985) \\
1420 & 80 $\pm$ 10 & Galt \& Kennedy (1968) \\
1720 & 66 $\pm$ 5 & Mantovani et al. (1985) \\
2695 & 51 $\pm$ 4 & Mantovani et al. (1985) \\
2700 & 42.2 $\pm$ 3.0 & Velusamy \& Kundu (1974) \\
4750 & 44 $\pm$ 3.3 &  Mantovani et al. (1985) \\
10200 & 18.8 $\pm$ 1.3  &  Sofue et al. (1983) \\
\enddata
\tablenotetext{a}{Reduced to Baars et al.'s (1977) scale for consistency.}
\end{deluxetable}

\end{document}